\documentclass[submission,copyright,creativecommons]{eptcs}
\providecommand{\event}{FMAS 2025} 

\usepackage{iftex}
\ifpdf
  \usepackage{underscore}
  \usepackage[T1]{fontenc}
\else
  \usepackage{breakurl}
\fi

\usepackage{tikz}

\title{Watchdogs and Oracles: Runtime Verification Meets Large Language Models for Autonomous Systems}
\author{Angelo Ferrando
\institute{University of Modena and Reggio Emilia\\ Modena, Italy}
\email{angelo.ferrando@unimore.it}
}
\def\titlerunning{Watchdogs and Oracles: RV Meets LLMs for Autonomous Systems}
\def\authorrunning{A. Ferrando}
\begin{document}
\maketitle

\begin{abstract}
Assuring the safety and trustworthiness of autonomous systems is particularly difficult when learning-enabled components and open environments are involved. Formal methods provide strong guarantees but depend on complete models and static assumptions. Runtime verification (RV) complements them by monitoring executions at run time and, in its predictive variants, by anticipating potential violations. Large language models (LLMs), meanwhile, excel at translating natural language into formal artefacts and recognising patterns in data, yet they remain error-prone and lack formal guarantees. This vision paper argues for a symbiotic integration of RV and LLMs. RV can serve as a guardrail for LLM-driven autonomy, while LLMs can extend RV by assisting specification capture, supporting anticipatory reasoning, and helping to handle uncertainty. We outline how this mutual reinforcement differs from existing surveys and roadmaps, discuss challenges and certification implications, and identify future research directions towards dependable autonomy.
\end{abstract}

\section{Introduction}

Autonomous systems are becoming increasingly pervasive, from self-driving vehicles navigating complex traffic environments, to collaborative industrial robots, to adaptive cyber-physical infrastructures. Their decision-making blends symbolic control with machine learning, and often relies on large-scale perception and reasoning components. While such systems promise unprecedented levels of autonomy and efficiency, they also pose acute challenges for safety, reliability, and accountability. Failures may lead not only to costly disruptions, but to physical harm and erosion of public trust.

Formal methods (FMs) have long provided mathematically rigorous tools to specify, model, and verify systems before deployment. However, these techniques typically assume complete and accurate models, and they encounter fundamental limitations when faced with opaque neural controllers, distributional shift, or highly dynamic and uncertain environments. To bridge this gap, \emph{runtime verification} (RV) \cite{DBLP:series/lncs/BartocciFFR18} has emerged as a complementary technique: instead of proving all properties ahead of time, RV monitors executions against temporal specifications and can detect or even enforce safety at run time \cite{DBLP:journals/fmsd/FalconeMFR11}. Recent advances in predictive and anticipatory RV extend this capability further by forecasting property violations before they occur, enabling proactive interventions \cite{DBLP:conf/nfm/ZhangLD12,DBLP:conf/cav/HiplerKLS24,DBLP:conf/cav/AngM24}.

In parallel, \emph{large language models} (LLMs) have transformed natural language processing and increasingly influence software engineering and formal methods. LLMs can translate informal requirements into formal artefacts, propose invariants, ranking functions, or even sketch verification algorithms \cite{DBLP:conf/vecos/CohenP23a,DBLP:conf/vecos/CohenP24,DBLP:journals/corr/abs-2502-18917,DBLP:conf/iclr/0001BN24}. By lowering the barrier to expressing formal properties, they democratise access to formal verification. Yet LLMs remain brittle: they hallucinate, misinterpret context, and provide no formal guarantees about their outputs.

This tension between the strengths and weaknesses of RV and LLMs motivates our vision. We argue that RV and LLMs can be integrated in a symbiotic way. RV can serve as a \emph{guardrail} for LLM-driven autonomy, ensuring safety and trustworthiness in execution. Conversely, LLMs can act as \emph{enablers} for RV, by generating specifications and monitors from natural language, predicting event patterns that extend beyond traditional logics, and assisting engineers in scaling monitoring infrastructures. Our focus in this paper is not on generic integration of LLMs and formal methods, which has already been surveyed elsewhere \cite{DBLP:journals/air/HuangRHJDWBMQZCZWXWFM24,DBLP:journals/corr/abs-2412-06512}, but specifically on the intersection of RV and LLMs in the context of autonomous systems.

We position this contribution as a vision paper: synthesising emerging evidence, highlighting lessons learned from recent experiments, and sketching promising avenues for future research. In particular, we emphasise (i) online and predictive monitoring of autonomous systems, (ii) LLM-assisted synthesis of specifications and monitors, and (iii) implications for certification and trustworthy deployment. By articulating this agenda, we aim to stimulate discussion in the formal methods community and to guide the development of practical, dependable solutions for learning-enabled autonomy.

To ground this agenda, Figure~\ref{fig:vision-architecture} depicts the interplay we envision between runtime verification and large language models. The diagram illustrates how the two technologies complement each other within a unified architecture. LLMs operate at three levels: at the front end, translating natural language into candidate specifications; alongside monitoring, contributing predictive continuations, gap-filling, and uncertainty handling; and within the system itself, where RV acts as a guardrail over their outputs. This dual perspective---LLMs as both enablers and subjects of monitoring---captures the essence of our vision: RV supplies formal assurance, while LLMs expand accessibility and foresight.
Importantly, in the role of subjects, LLMs are treated as opaque components whose outputs are supervised by RV monitors. Even though their internal reasoning cannot be verified, RV ensures that their observable behaviour complies with safety constraints. The model illustrated here will serve as a reference throughout the remainder of the paper.

\tikzset{every picture/.style={line width=0.75pt}} 

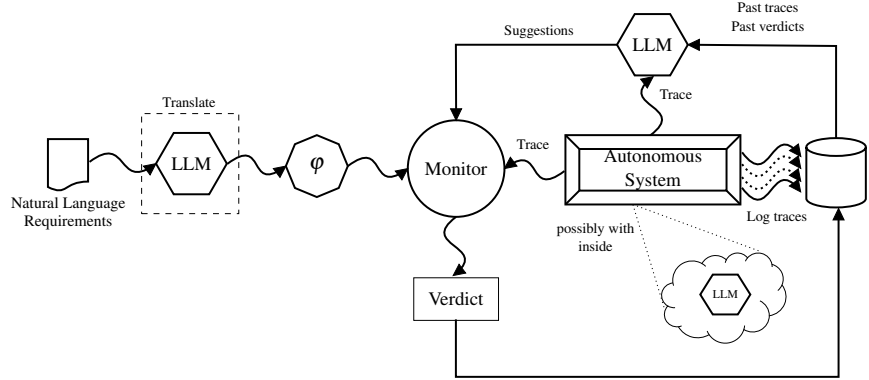
\begin{figure}
    \centering
    \scalebox{0.5}{
\begin{tikzpicture}[x=0.75pt,y=0.75pt,yscale=-1,xscale=1]

\draw  [line width=1.5]  (430,178.75) .. controls (430,151.83) and (451.83,130) .. (478.75,130) .. controls (505.67,130) and (527.5,151.83) .. (527.5,178.75) .. controls (527.5,205.67) and (505.67,227.5) .. (478.75,227.5) .. controls (451.83,227.5) and (430,205.67) .. (430,178.75) -- cycle ;
\draw  [line width=1.5]  (67,149) -- (107.5,149) -- (107.5,191.9) .. controls (82.19,191.9) and (87.25,207.37) .. (67,197.36) -- cycle ;
\draw [line width=1.5]    (109,172) .. controls (148,142.75) and (136.62,200.72) .. (173.1,175.83) ;
\draw [shift={(176,173.75)}, rotate = 143.13] [fill={rgb, 255:red, 0; green, 0; blue, 0 }  ][line width=0.08]  [draw opacity=0] (11.61,-5.58) -- (0,0) -- (11.61,5.58) -- cycle    ;
\draw [line width=1.5]    (248,174) .. controls (287,144.75) and (268.01,203.2) .. (304.11,178.33) ;
\draw [shift={(307,176.25)}, rotate = 143.13] [fill={rgb, 255:red, 0; green, 0; blue, 0 }  ][line width=0.08]  [draw opacity=0] (11.61,-5.58) -- (0,0) -- (11.61,5.58) -- cycle    ;
\draw  [dash pattern={on 4.5pt off 4.5pt}] (161.5,123.25) -- (259.5,123.25) -- (259.5,224.25) -- (161.5,224.25) -- cycle ;
\draw  [line width=1.5]  (247.5,173.75) -- (229.63,204.71) -- (193.88,204.71) -- (176,173.75) -- (193.88,142.79) -- (229.63,142.79) -- cycle ;
\draw  [line width=1.5]  (369.5,176.25) -- (360.35,198.35) -- (338.25,207.5) -- (316.15,198.35) -- (307,176.25) -- (316.15,154.15) -- (338.25,145) -- (360.35,154.15) -- cycle ;
\draw [line width=1.5]    (369.5,176.25) .. controls (408.5,147) and (390.94,205.68) .. (427.11,180.83) ;
\draw [shift={(430,178.75)}, rotate = 143.13] [fill={rgb, 255:red, 0; green, 0; blue, 0 }  ][line width=0.08]  [draw opacity=0] (11.61,-5.58) -- (0,0) -- (11.61,5.58) -- cycle    ;
\draw  [line width=1.5]  (589,145) -- (765.5,145) -- (765.5,215) -- (589,215) -- cycle ; \draw  [line width=1.5]  (603,159) -- (751.5,159) -- (751.5,201) -- (603,201) -- cycle ; \draw  [line width=1.5]  (589,145) -- (603,159) ; \draw  [line width=1.5]  (765.5,145) -- (751.5,159) ; \draw  [line width=1.5]  (765.5,215) -- (751.5,201) ; \draw  [line width=1.5]  (589,215) -- (603,201) ;
\draw [line width=1.5]    (530.94,176.29) .. controls (566.38,152.42) and (549.2,210.35) .. (588,181.25) ;
\draw [shift={(527.5,178.75)}, rotate = 323.13] [fill={rgb, 255:red, 0; green, 0; blue, 0 }  ][line width=0.08]  [draw opacity=0] (11.61,-5.58) -- (0,0) -- (11.61,5.58) -- cycle    ;
\draw  [line width=1.5]  (712.5,53.75) -- (694.63,84.71) -- (658.88,84.71) -- (641,53.75) -- (658.88,22.79) -- (694.63,22.79) -- cycle ;
\draw [line width=1.5]    (767.21,162.34) .. controls (796.3,201.46) and (796.6,130.19) .. (824.26,163.78) ;
\draw [shift={(826.44,166.58)}, rotate = 233.36] [fill={rgb, 255:red, 0; green, 0; blue, 0 }  ][line width=0.08]  [draw opacity=0] (11.61,-5.58) -- (0,0) -- (11.61,5.58) -- cycle    ;
\draw [line width=1.5]    (767.06,198.09) .. controls (796.15,237.21) and (796.46,165.94) .. (824.11,199.53) ;
\draw [shift={(826.29,202.33)}, rotate = 233.36] [fill={rgb, 255:red, 0; green, 0; blue, 0 }  ][line width=0.08]  [draw opacity=0] (11.61,-5.58) -- (0,0) -- (11.61,5.58) -- cycle    ;
\draw [line width=1.5]  [dash pattern={on 1.69pt off 2.76pt}]  (767.16,174.34) .. controls (796.25,213.46) and (796.55,142.19) .. (824.21,175.78) ;
\draw [shift={(826.39,178.58)}, rotate = 233.36] [fill={rgb, 255:red, 0; green, 0; blue, 0 }  ][line width=0.08]  [draw opacity=0] (11.61,-5.58) -- (0,0) -- (11.61,5.58) -- cycle    ;
\draw [line width=1.5]  [dash pattern={on 1.69pt off 2.76pt}]  (767.1,187.34) .. controls (796.2,226.46) and (796.5,155.19) .. (824.16,188.78) ;
\draw [shift={(826.34,191.58)}, rotate = 233.36] [fill={rgb, 255:red, 0; green, 0; blue, 0 }  ][line width=0.08]  [draw opacity=0] (11.61,-5.58) -- (0,0) -- (11.61,5.58) -- cycle    ;
\draw  [line width=1.5]  (892,157) -- (892,209) .. controls (892,213.97) and (878.57,218) .. (862,218) .. controls (845.43,218) and (832,213.97) .. (832,209) -- (832,157) .. controls (832,152.03) and (845.43,148) .. (862,148) .. controls (878.57,148) and (892,152.03) .. (892,157) .. controls (892,161.97) and (878.57,166) .. (862,166) .. controls (845.43,166) and (832,161.97) .. (832,157) ;
\draw [line width=1.5]    (674.5,145.25) .. controls (713.5,116) and (641.28,115.98) .. (674.72,88.19) ;
\draw [shift={(677.5,86)}, rotate = 143.13] [fill={rgb, 255:red, 0; green, 0; blue, 0 }  ][line width=0.08]  [draw opacity=0] (11.61,-5.58) -- (0,0) -- (11.61,5.58) -- cycle    ;
\draw [line width=1.5]    (862,148) -- (862,55) -- (716.5,53.78) ;
\draw [shift={(712.5,53.75)}, rotate = 0.48] [fill={rgb, 255:red, 0; green, 0; blue, 0 }  ][line width=0.08]  [draw opacity=0] (11.61,-5.58) -- (0,0) -- (11.61,5.58) -- cycle    ;
\draw [line width=1.5]    (641,53.75) -- (478.5,53.75) -- (478.74,126) ;
\draw [shift={(478.75,130)}, rotate = 269.81] [fill={rgb, 255:red, 0; green, 0; blue, 0 }  ][line width=0.08]  [draw opacity=0] (11.61,-5.58) -- (0,0) -- (11.61,5.58) -- cycle    ;
\draw [line width=1.5]    (479.04,284.13) .. controls (511.24,256.77) and (439.95,256.6) .. (478.75,227.5) ;
\draw [shift={(475.75,286.75)}, rotate = 323.13] [fill={rgb, 255:red, 0; green, 0; blue, 0 }  ][line width=0.08]  [draw opacity=0] (11.61,-5.58) -- (0,0) -- (11.61,5.58) -- cycle    ;
\draw   (436,289) -- (521.5,289) -- (521.5,332) -- (436,332) -- cycle ;
\draw  [line width=1.5]  (775.5,305.89) -- (762.88,326) -- (737.63,326) -- (725,305.89) -- (737.63,285.79) -- (762.88,285.79) -- cycle ;

\draw   (697.42,293.96) .. controls (696.41,286.62) and (699.73,279.36) .. (705.97,275.25) .. controls (712.21,271.15) and (720.27,270.91) .. (726.74,274.66) .. controls (729.03,270.39) and (733.22,267.45) .. (738.05,266.71) .. controls (742.88,265.98) and (747.77,267.54) .. (751.26,270.93) .. controls (753.21,267.06) and (757.04,264.46) .. (761.39,264.06) .. controls (765.75,263.65) and (770.01,265.49) .. (772.66,268.93) .. controls (776.19,264.83) and (781.8,263.1) .. (787.07,264.5) .. controls (792.33,265.89) and (796.31,270.15) .. (797.28,275.45) .. controls (801.6,276.61) and (805.2,279.57) .. (807.14,283.56) .. controls (809.09,287.55) and (809.2,292.18) .. (807.43,296.26) .. controls (811.69,301.73) and (812.68,309.02) .. (810.05,315.41) .. controls (807.41,321.8) and (801.54,326.32) .. (794.63,327.3) .. controls (794.58,333.3) and (791.25,338.8) .. (785.92,341.68) .. controls (780.6,344.57) and (774.11,344.39) .. (768.95,341.22) .. controls (766.76,348.4) and (760.58,353.68) .. (753.09,354.79) .. controls (745.59,355.89) and (738.13,352.62) .. (733.92,346.38) .. controls (728.76,349.45) and (722.57,350.34) .. (716.74,348.84) .. controls (710.91,347.33) and (705.94,343.57) .. (702.95,338.39) .. controls (697.67,339) and (692.57,336.3) .. (690.18,331.63) .. controls (687.79,326.96) and (688.61,321.32) .. (692.24,317.5) .. controls (687.53,314.76) and (685.13,309.34) .. (686.29,304.04) .. controls (687.44,298.75) and (691.89,294.8) .. (697.31,294.24) ; \draw   (692.24,317.5) .. controls (694.46,318.79) and (697.02,319.38) .. (699.59,319.18)(702.95,338.39) .. controls (704.05,338.26) and (705.13,337.99) .. (706.16,337.59)(733.92,346.38) .. controls (733.14,345.23) and (732.49,344) .. (731.98,342.71)(768.95,341.22) .. controls (769.35,339.91) and (769.61,338.56) .. (769.73,337.2)(794.62,327.3) .. controls (794.68,320.92) and (791.01,315.07) .. (785.19,312.28)(807.43,296.26) .. controls (806.49,298.43) and (805.05,300.36) .. (803.23,301.89)(797.28,275.45) .. controls (797.44,276.32) and (797.51,277.21) .. (797.5,278.11)(772.66,268.93) .. controls (771.78,269.95) and (771.06,271.09) .. (770.51,272.32)(751.26,270.93) .. controls (750.79,271.86) and (750.44,272.84) .. (750.21,273.86)(726.74,274.66) .. controls (728.11,275.45) and (729.37,276.4) .. (730.51,277.5)(697.42,293.96) .. controls (697.56,294.97) and (697.78,295.97) .. (698.08,296.95) ;
\draw  [dash pattern={on 0.84pt off 2.51pt}]  (657,215) -- (688.5,319) ;
\draw  [dash pattern={on 0.84pt off 2.51pt}]  (657,215) -- (784.5,262) ;
\draw [line width=1.5]    (478.5,331) -- (478.5,389) -- (864.5,389) -- (864.5,223) ;
\draw [shift={(864.5,219)}, rotate = 90] [fill={rgb, 255:red, 0; green, 0; blue, 0 }  ][line width=0.08]  [draw opacity=0] (11.61,-5.58) -- (0,0) -- (11.61,5.58) -- cycle    ;

\draw (478.75,178.75) node  [font=\Large] [align=left] {Monitor};
\draw (87.5,205.9) node [anchor=north] [inner sep=0.75pt]  [font=\large] [align=left] {\begin{minipage}[lt]{100.08pt}\setlength\topsep{0pt}
\begin{center}
Natural Language\\Requirements
\end{center}

\end{minipage}};
\draw (211.75,173.75) node  [font=\Large] [align=left] {LLM};
\draw (209.5,104.9) node [anchor=north] [inner sep=0.75pt]  [font=\normalsize] [align=left] {\begin{minipage}[lt]{44.88pt}\setlength\topsep{0pt}
\begin{center}
Translate
\end{center}

\end{minipage}};
\draw (338.25,176.25) node  [font=\LARGE] [align=left] {\begin{minipage}[lt]{16.34pt}\setlength\topsep{0pt}
\begin{center}
$\displaystyle \varphi $
\end{center}

\end{minipage}};
\draw (677.25,180) node  [font=\Large] [align=center] {Autonomous \\System};
\draw (556.5,147.9) node [anchor=north] [inner sep=0.75pt]  [font=\normalsize] [align=left] {\begin{minipage}[lt]{28.42pt}\setlength\topsep{0pt}
\begin{center}
Trace
\end{center}

\end{minipage}};
\draw (676.75,53.75) node  [font=\Large] [align=left] {LLM};
\draw (802.5,219.9) node [anchor=north] [inner sep=0.75pt]  [font=\normalsize] [align=left] {\begin{minipage}[lt]{50.38pt}\setlength\topsep{0pt}
\begin{center}
Log traces
\end{center}

\end{minipage}};
\draw (700.5,96.9) node [anchor=north] [inner sep=0.75pt]  [font=\normalsize] [align=left] {\begin{minipage}[lt]{28.42pt}\setlength\topsep{0pt}
\begin{center}
Trace
\end{center}

\end{minipage}};
\draw (793.5,9.9) node [anchor=north] [inner sep=0.75pt]  [font=\normalsize] [align=left] {\begin{minipage}[lt]{53.75pt}\setlength\topsep{0pt}
\begin{center}
Past traces
\end{center}

\end{minipage}};
\draw (562.5,32.9) node [anchor=north] [inner sep=0.75pt]  [font=\normalsize] [align=left] {\begin{minipage}[lt]{58.89pt}\setlength\topsep{0pt}
\begin{center}
Suggestions
\end{center}

\end{minipage}};
\draw (478.75,310.5) node  [font=\Large] [align=left] {Verdict};
\draw (750.25,305.89) node  [font=\Large] [align=left] {{\footnotesize LLM}};
\draw (619.5,231.9) node [anchor=north] [inner sep=0.75pt]  [font=\normalsize] [align=left] {\begin{minipage}[lt]{60.57pt}\setlength\topsep{0pt}
\begin{center}
possibly with\\ inside
\end{center}

\end{minipage}};
\draw (793.5,27.9) node [anchor=north] [inner sep=0.75pt]  [font=\normalsize] [align=left] {\begin{minipage}[lt]{61.12pt}\setlength\topsep{0pt}
\begin{center}
Past verdicts
\end{center}

\end{minipage}};

\end{tikzpicture}
}
    \caption{Envisioned architecture for RV+LLM integration. LLMs contribute as (i) enablers, translating informal requirements into candidate specifications, (ii) collaborators, providing predictive continuations to support anticipatory RV, and (iii) subjects, whose outputs are supervised by RV monitors to ensure safety. This dual role illustrates the bidirectional nature of our vision.}
    \label{fig:vision-architecture}
\end{figure}

\section{Related Work}

The integration of LLMs and formal methods has attracted growing attention. 
Huang et al.\ present a comprehensive survey of techniques for assessing the safety and trustworthiness of LLMs, 
covering static analysis, dynamic validation, testing frameworks, and monitoring mechanisms 
\cite{DBLP:journals/air/HuangRHJDWBMQZCZWXWFM24}. 
Their scope is intentionally broad: runtime monitoring appears as one tool among many, alongside interpretability, robustness evaluation, and adversarial testing.  

Complementing this, Zhang et al.\ propose a roadmap for combining formal methods and LLMs to build trustworthy AI agents \cite{DBLP:journals/corr/abs-2412-06512}. 
A more recent survey specifically examines the use of LLMs for specification capture and analysis \cite{DBLP:journals/corr/abs-2507-14330}. 
Together, these surveys discuss challenges across deductive verification, theorem proving, model checking, and certification, 
with a largely technology-agnostic and domain-independent perspective. 
By contrast, our contribution is narrower but deeper: we focus specifically on runtime verification in the context of autonomous systems, 
emphasising predictive monitoring and imperfect-information settings.

More specialised efforts have examined concrete intersections of LLMs and runtime verification. 
Cohen and Peled show how LLMs can aid monitor construction by translating natural-language requirements into formal specifications 
\cite{DBLP:conf/vecos/CohenP24}. 
Such work demonstrates how LLMs can reduce the entry barrier to RV, but it remains focused on static monitor synthesis rather than online monitoring.  

In parallel, predictive runtime verification research investigates how monitors can conclude verdicts with \emph{less observation} than classical semantics would require, 
by exploiting system models or reasoning over partial trace prefixes 
\cite{DBLP:conf/cav/HiplerKLS24,DBLP:conf/cav/AngM24,DBLP:conf/nfm/ZhangLD12}. 
This anticipatory capability is critical in autonomous domains, where delayed detection can lead to unsafe behaviour. 
Here we see a natural role for LLMs: they could help infer approximate system models from informal descriptions, 
or reduce the set of future events that must still be considered (for instance, by recognising that certain actions will not reoccur). 
Integrating such anticipatory monitoring with LLM-assisted modelling could yield monitors that are both timely and practical in complex, data-rich environments. 

A different strand of work treats the LLM itself as the subject of monitoring. Pro2Guard, for instance, abstracts the behaviour of an LLM agent into a Discrete-Time Markov Chain (DTMC) and applies probabilistic model checking to anticipate unsafe outcomes \cite{wang2025pro2guardproactiveruntimeenforcement}. In contrast, Statistical Runtime Verification for LLMs adopts a black-box view, using robustness estimation to bound confidence under perturbations \cite{Levy2025StatisticalRV}. Although these approaches differ in technique, they share the same perspective: RV acts as an external safeguard around the LLM, but the LLM does not contribute to the monitoring process.
Our vision inverts this relationship. We propose that LLMs should not only be monitored but also play an active role in enhancing RV itself---supporting anticipatory reasoning and helping monitors cope with imperfect information.

In summary, existing surveys and roadmaps consider the broad landscape of LLMs and formal methods, 
while specific contributions address either monitor synthesis or the monitoring of LLMs as opaque systems. 
By contrast, our vision focuses on the synergy of LLMs and runtime verification within \emph{autonomous systems}. 
Domains such as medical robotics, autonomous driving, and adaptive industrial systems pose distinctive challenges of real-time monitoring, incomplete information, and safety-critical decision making. 
We argue that combining RV’s formal rigour and online guarantees with the adaptive and predictive capabilities of LLMs 
offers a promising path toward trustworthy, anticipatory supervision for autonomy---distinct from broader, technology-agnostic surveys and roadmaps in the literature.

\section{LLMs in Support of Runtime Verification}

One clear opportunity for synergy is using LLMs to lower the barrier to RV. In industrial practice, requirements are overwhelmingly expressed in natural language, and translating them into temporal logic is labour-intensive and error-prone. LLMs can assist here: Cohen and Peled demonstrated that monitors can be synthesised from plain-English descriptions via an iterative prompt-and-check loop \cite{DBLP:conf/vecos/CohenP23a}, while follow-up work showed how LLMs may even propose whole verification algorithms in pseudocode, to be validated and refined by experts \cite{DBLP:conf/vecos/CohenP24}. Similarly, tools such as ClassInvGen \cite{DBLP:journals/corr/abs-2502-18917} and Lemur \cite{DBLP:conf/iclr/0001BN24} exploit LLMs to suggest invariants or ranking functions, with static analysers providing rigorous validation.  


Where LLMs may offer a more transformative role is in \emph{predictive RV}. Unlike classical monitoring---checking satisfaction on an observed trace---predictive RV ventures into the realm of uncertainty: anticipating likely future events and estimating violation probabilities. The approach by Zhang, Leucker and Dong \cite{DBLP:conf/nfm/ZhangLD12} formalises predictive semantics for runtime verification, enabling monitors to look ahead beyond the current execution prefix. Relatedly, work on imperfect-information monitoring \cite{DBLP:conf/sefm/FerrandoM22} addresses how to reason when only partial observations of the system are available. LLMs could support such approaches by learning from traces to propose plausible continuations (filling gaps) or high-level predictions that, when fed into formally synthesised monitors, enable earlier and more informed interventions.

LLMs could significantly extend these predictive capabilities. By learning from historical traces or domain-specific corpora, they can recognise subtle patterns of behaviour and propose likely continuations of execution. Coupled with formal monitors, such insights enable anticipatory interventions: not merely flagging a violation once it occurs, but estimating its probability ahead of time and prompting earlier corrective action. For instance, an LLM trained on driving logs might infer that after observing a pedestrian near a crossing, a braking event is highly likely. Feeding this continuation into a predictive RV monitor allows the system to assess the probability of a safety violation sooner than observation alone would permit. Such foresight is particularly valuable in safety-critical domains where avoiding unsafe conditions requires anticipation as much as detection.

Thus, while we maintain that \emph{monitor synthesis and enforcement must remain formal}, LLMs offer a promising role in the uncertain terrain of predictive RV---extrapolating, filling gaps, and proposing continuations where purely logic-based approaches struggle. When such predictions are unreliable or misleading, the RV framework can fall back to a conservative mode of operation: discarding unverifiable suggestions and reverting to classical monitoring semantics. This safeguard ensures that correctness is never compromised, even when LLM assistance fails. Exploring this balance between formal guarantees and predictive support represents, in our view, one of the most exciting directions for future research.

\section{Runtime Verification to Guard LLMs}

The complementary direction treats LLM-driven components themselves as subjects of monitoring and enforcement. Unlike traditional software modules, LLMs are stochastic, non-deterministic, and continually updated. Their internal reasoning processes are opaque, making pre-deployment formal verification impractical. This opacity and adaptivity make LLMs natural candidates for oversight by RV, which can provide on-line assurances even when static guarantees are unattainable.  

Recent approaches illustrate this trend. Zhang et al.\ propose \emph{RvLLM} \cite{DBLP:journals/corr/abs-2505-18585}, which augments LLMs with domain knowledge and enforces correctness by checking responses against safety constraints at run time, retrying or abstaining upon violations. Levy et al.\ introduce a \emph{statistical} RV framework that interprets LLM confidence scores through probabilistic monitoring, providing robustness estimates when deterministic guarantees are out of reach \cite{Levy2025StatisticalRV}. These efforts highlight two central insights: (i) RV can act as a lightweight safety layer around untrusted models, and (ii) monitoring must be adapted to cope with inherently probabilistic behaviour.  

A particularly promising avenue is the integration of \emph{predictive RV}. Anticipatory monitors can compute lookahead satisfaction probabilities and pre-empt unsafe actions \cite{DBLP:conf/cav/HiplerKLS24}, while predictive prefixes allow violations of temporal logic properties to be detected earlier \cite{DBLP:conf/cav/AngM24}. When applied to LLMs, these techniques could detect signs of drift or hallucination before erroneous outputs are delivered, e.g.\ flagging an unsafe instruction in dialogue before it is executed by a robotic agent.  

The challenge becomes even sharper in settings with \emph{distribution shift} or \emph{incomplete information}, which are common in LLM deployments. Distributionally robust and conformal prediction monitors \cite{DBLP:journals/corr/abs-2504-02964,DBLP:conf/allerton/LindemannQDP23} quantify uncertainty when the data distribution changes, a scenario directly relevant to LLMs exposed to novel prompts or adversarial contexts. Similarly, techniques for runtime verification with imperfect information \cite{DBLP:conf/sefm/FerrandoM22} suggest that monitoring LLM behaviour under partial observability is not only possible but essential: the full reasoning state of an LLM is inaccessible, and only outputs (and limited internal signals such as probabilities) can be observed.  

Taken together, these approaches indicate that RV can serve as a multi-layered guard for LLMs:  
\begin{itemize}
  \item \textbf{Syntactic monitoring}, filtering outputs for forbidden tokens or patterns.  
  \item \textbf{Semantic monitoring}, enforcing domain-specific temporal or safety constraints.  
  \item \textbf{Predictive monitoring}, anticipating unsafe continuations, hallucinations, or distribution shifts before they manifest.  
\end{itemize}

This layered perspective also connects to certification. Since LLM internals cannot be fully verified, regulators may instead require runtime assurance frameworks that combine statistical confidence measures with formal monitors. We envision RV not merely as a reactive guardrail but as an anticipatory partner, enabling dependable use of LLMs within safety-critical autonomous systems.

To situate our vision within the broader landscape, Table~\ref{tab:landscape} contrasts three existing lines of work---LLM-assisted specification, monitoring of LLMs, and predictive RV---with the perspective advanced in this paper. While Figure \ref{fig:vision-architecture} illustrated how RV and LLMs can interact within a unified architecture, the table highlights how prior research has addressed only fragments of this picture. Our contribution differs in combining these strands: positioning LLMs not just as objects of monitoring or front-end translators, but as active components that can enhance predictive and uncertainty-aware runtime verification.

\begin{table}[ht]
\centering
\caption{Landscape of RV+LLM research and this vision.}
\label{tab:landscape}
\scriptsize
\begin{tabular}{p{3.2cm}p{3.2cm}p{3.2cm}p{3.2cm}}
\hline
\textbf{Line of work} & \textbf{Role of LLM} & \textbf{Role of RV} & \textbf{Limitation} \\
\hline
LLM $\rightarrow$ spec translation 
(e.g.\ Cohen \& Peled) & NL $\rightarrow$ logic/specs; front-end assistant & Monitor synthesis; offline validation & Static; no anticipation; requires human vetting \\
\hline
Monitoring LLMs 
(e.g.\ Pro2Guard; Statistical RV) & LLM = monitored object; black-box outputs & Guardrail at runtime; enforcement / risk checks & LLM does not assist monitor; external, not bidirectional \\
\hline
Predictive RV (formal) 
(e.g.\ Hipler et al.; Ang \& Mathur; Zhang–Leucker–Dong) & -- & Anticipation via models/prefixes; early verdicts with less observation & No LLM-aided anticipation; no uncertainty resolution \\
\hline
\textbf{This vision paper} & \textbf{Active aide: prediction, modelling, gap-filling} & \textbf{Formal guardrail + anticipatory partner} & \textbf{Validation and certification challenges remain} \\
\hline
\end{tabular}

\vspace{3pt}
\small
\emph{Citations (not in cells):}
LLM$\rightarrow$spec: \cite{DBLP:conf/vecos/CohenP24,DBLP:conf/vecos/CohenP23a}.
Monitoring LLMs: \cite{wang2025pro2guardproactiveruntimeenforcement,Levy2025StatisticalRV}.
Predictive RV (formal): \cite{DBLP:conf/cav/HiplerKLS24,DBLP:conf/cav/AngM24,DBLP:conf/nfm/ZhangLD12}.
\end{table}

\section{Challenges and Certification Implications}

The integration of RV and LLMs holds great promise, but also exposes a set of difficult open problems:

\begin{itemize}
    \item \textbf{Trust.} LLM-generated artefacts---whether properties, invariants, or predictions---cannot be accepted at face value. Without formal validation, subtle flaws may undermine the very safety guarantees that RV seeks to provide. 

    \item \textbf{Efficiency.} Predictive monitors must operate within strict real-time budgets, especially in domains such as driving or medical robotics where delays are unacceptable. LLM queries may exceed these constraints, making lightweight distilled models or offline predictors necessary. Determining the boundary of feasible deployment remains an open question.

    \item \textbf{Human--AI collaboration.} Interfaces and workflows must be designed so that LLM-assisted RV remains intelligible, supporting rather than obscuring engineers’ reasoning about system safety. 

    \item \textbf{Robustness.} Both monitors and LLMs must continue to function under distribution shift or even adversarial manipulation. 

    \item \textbf{Ethics and accountability.} When system behaviour results from LLM suggestions filtered through monitors, the locus of responsibility must remain transparent to maintain trust and assign liability.
\end{itemize}

These challenges feed directly into the problem of \emph{certification}. 
Standards such as ISO 26262 for automotive functional safety \cite{ISO26262}, DO-178C for avionics software certification \cite{DBLP:conf/adaEurope/Daniels11}, and the forthcoming EU AI Act regulating trustworthy AI \cite{EUAIAct} demand structured evidence of safety, traceability, and compliance.

RV+LLM pipelines could provide certification evidence in novel ways.
\emph{LLM-assisted specification capture} can establish transparent links from informal requirements to executable monitors, improving traceability.
\emph{Predictive RV enhanced by LLM-based modelling or forecasting} can generate runtime assurance artefacts that complement offline analyses even in real-time environments, offering regulators a dynamic view of system safety.
Finally, \emph{monitor-assisted explanations of rejected LLM outputs} can increase auditability, clarifying to both engineers and auditors why certain behaviours were deemed unsafe.
For this potential to materialise, however, certification frameworks themselves must evolve to recognise dynamic, runtime evidence as a legitimate complement to traditional verification artefacts.

Notably, existing predictive-RV frameworks rely on formal abstractions to anticipate violations, but none exploit LLMs to assist anticipation or to reduce the set of future observations needed for early conclusions. 
Similarly, imperfect-information monitoring has been studied through logical and automata-theoretic techniques, but not with LLMs acting as uncertainty resolvers. 
This gap underscores the novelty of our proposal: to treat LLMs not merely as translators of specifications, but as \emph{active components of runtime monitors}, augmenting anticipatory reasoning and strengthening monitoring under uncertainty.

\section{Conclusions and Future Work}

Addressing the challenges of integrating RV and LLMs calls for progress on several interconnected fronts.
A first priority is \emph{hybrid predictive RV}, where LLM-based sequence prediction is combined with formally synthesised monitors to deliver earlier verdicts than traditional logics permit.
A second is \emph{domain-specific adaptation}: tailoring LLMs with corpora of safety standards and certification artefacts can constrain their outputs, yielding more precise and auditable specifications.
Equally important are \emph{interactive toolchains} that integrate generation, human refinement, and formal validation into a single loop, making monitor construction both collaborative and transparent.
Finally, certification practices must evolve to embrace these innovations, formally recognising runtime assurance evidence from RV+LLM systems as part of assurance cases.

Our focus here is on RV, yet similar bidirectional opportunities may exist for other formal techniques: LLMs could assist in invariant discovery for model checking or proof search in theorem proving, while those methods in turn could provide stronger guarantees for LLM outputs.
By contrast, much current research frames RV as a safeguard \emph{around} LLMs. Our perspective inverts this relationship: LLMs themselves can act as \emph{enablers} of novel forms of RV, supporting anticipatory reasoning and helping to resolve uncertainty. This inversion opens a distinct research trajectory that complements existing monitoring approaches.

Taken together, these priorities suggest a compelling agenda.
LLMs show promising ability to assist in translating natural language into formal artefacts, but remain imperfect and error-prone; RV provides enforcement and formal assurance. Their combination can lower the barrier to formalising requirements, enable anticipatory supervision of learning-enabled components, and enhance the trustworthiness of autonomous systems.
Unlike broader surveys and roadmaps \cite{DBLP:journals/corr/abs-2412-06512,DBLP:journals/air/HuangRHJDWBMQZCZWXWFM24}, our focus is specifically on real-time monitoring, predictive semantics, and domain-specific deployment.
Realising this vision will require advances in predictive reasoning, robust integration with human workflows, and stronger alignment with certification practices.
Yet the potential benefits---dependable autonomy in medicine, transport, and generally in safety-critical infrastructures---make this a timely and urgent research direction.

\bibliographystyle{eptcs}
\bibliography{main}

\begin{thebibliography}{10}
\providecommand{\bibitemdeclare}[2]{}
\providecommand{\surnamestart}{}
\providecommand{\surnameend}{}
\providecommand{\urlprefix}{Available at }
\providecommand{\url}[1]{\texttt{#1}}
\providecommand{\href}[2]{\texttt{#2}}
\providecommand{\urlalt}[2]{\href{#1}{#2}}
\providecommand{\doi}[1]{doi:\urlalt{https://doi.org/#1}{#1}}
\providecommand{\eprint}[1]{arXiv:\urlalt{https://arxiv.org/abs/#1}{#1}}
\providecommand{\bibinfo}[2]{#2}

\bibitemdeclare{inproceedings}{DBLP:conf/cav/AngM24}
\bibitem{DBLP:conf/cav/AngM24}
\bibinfo{author}{Zhendong \surnamestart Ang\surnameend} \& \bibinfo{author}{Umang \surnamestart Mathur\surnameend} (\bibinfo{year}{2024}): \emph{\bibinfo{title}{Predictive Monitoring with Strong Trace Prefixes}}.
\newblock In \bibinfo{editor}{Arie \surnamestart Gurfinkel\surnameend} \& \bibinfo{editor}{Vijay \surnamestart Ganesh\surnameend}, editors: {\slshape \bibinfo{booktitle}{Computer Aided Verification - 36th International Conference, {CAV} 2024, Montreal, QC, Canada, July 24-27, 2024, Proceedings, Part {II}}}, {\slshape \bibinfo{series}{Lecture Notes in Computer Science}} \bibinfo{volume}{14682}, \bibinfo{publisher}{Springer}, pp. \bibinfo{pages}{182--204}, \doi{10.1007/978-3-031-65630-9\_9}.

\bibitemdeclare{incollection}{DBLP:series/lncs/BartocciFFR18}
\bibitem{DBLP:series/lncs/BartocciFFR18}
\bibinfo{author}{Ezio \surnamestart Bartocci\surnameend}, \bibinfo{author}{Yli{\`{e}}s \surnamestart Falcone\surnameend}, \bibinfo{author}{Adrian \surnamestart Francalanza\surnameend} \& \bibinfo{author}{Giles \surnamestart Reger\surnameend} (\bibinfo{year}{2018}): \emph{\bibinfo{title}{Introduction to Runtime Verification}}.
\newblock In \bibinfo{editor}{Ezio \surnamestart Bartocci\surnameend} \& \bibinfo{editor}{Yli{\`{e}}s \surnamestart Falcone\surnameend}, editors: {\slshape \bibinfo{booktitle}{Lectures on Runtime Verification - Introductory and Advanced Topics}}, {\slshape \bibinfo{series}{Lecture Notes in Computer Science}} \bibinfo{volume}{10457}, \bibinfo{publisher}{Springer}, pp. \bibinfo{pages}{1--33}, \doi{10.1007/978-3-319-75632-5\_1}.

\bibitemdeclare{article}{DBLP:journals/corr/abs-2507-14330}
\bibitem{DBLP:journals/corr/abs-2507-14330}
\bibinfo{author}{Arshad \surnamestart Beg\surnameend}, \bibinfo{author}{Diarmuid~P. \surnamestart O'Donoghue\surnameend} \& \bibinfo{author}{Rosemary \surnamestart Monahan\surnameend} (\bibinfo{year}{2025}): \emph{\bibinfo{title}{Leveraging LLMs for Formal Software Requirements - Challenges and Prospects}}.
\newblock {\slshape \bibinfo{journal}{CoRR}} \bibinfo{volume}{abs/2507.14330}, \doi{10.48550/ARXIV.2507.14330}.
\newblock \eprint{2507.14330}.

\bibitemdeclare{inproceedings}{DBLP:conf/vecos/CohenP23a}
\bibitem{DBLP:conf/vecos/CohenP23a}
\bibinfo{author}{Itay \surnamestart Cohen\surnameend} \& \bibinfo{author}{Doron \surnamestart Peled\surnameend} (\bibinfo{year}{2023}): \emph{\bibinfo{title}{End-to-End {AI} Generated Runtime Verification from Natural Language Specification}}.
\newblock In \bibinfo{editor}{Bernhard \surnamestart Steffen\surnameend}, editor: {\slshape \bibinfo{booktitle}{Bridging the Gap Between {AI} and Reality - First International Conference, AISoLA 2023, Crete, Greece, October 23-28, 2023, Selected Papers}}, {\slshape \bibinfo{series}{Lecture Notes in Computer Science}} \bibinfo{volume}{14129}, \bibinfo{publisher}{Springer}, pp. \bibinfo{pages}{362--384}, \doi{10.1007/978-3-031-73741-1\_23}.

\bibitemdeclare{inproceedings}{DBLP:conf/vecos/CohenP24}
\bibitem{DBLP:conf/vecos/CohenP24}
\bibinfo{author}{Itay \surnamestart Cohen\surnameend} \& \bibinfo{author}{Doron \surnamestart Peled\surnameend} (\bibinfo{year}{2024}): \emph{\bibinfo{title}{LLM-Based Scheme for Synthesis of Formal Verification Algorithms}}.
\newblock In \bibinfo{editor}{Bernhard \surnamestart Steffen\surnameend}, editor: {\slshape \bibinfo{booktitle}{Bridging the Gap Between {AI} and Reality - Second International Conference, AISoLA 2024, Crete, Greece, October 30 - November 3, 2024, Proceedings}}, {\slshape \bibinfo{series}{Lecture Notes in Computer Science}} \bibinfo{volume}{15217}, \bibinfo{publisher}{Springer}, pp. \bibinfo{pages}{167--182}, \doi{10.1007/978-3-031-75434-0\_11}.

\bibitemdeclare{inproceedings}{DBLP:conf/adaEurope/Daniels11}
\bibitem{DBLP:conf/adaEurope/Daniels11}
\bibinfo{author}{Dewi \surnamestart Daniels\surnameend} (\bibinfo{year}{2011}): \emph{\bibinfo{title}{Position Paper: {DO-178C/ED-12C} and Object-Orientation for Critical Systems}}.
\newblock In \bibinfo{editor}{Alexander~B. \surnamestart Romanovsky\surnameend} \& \bibinfo{editor}{Tullio \surnamestart Vardanega\surnameend}, editors: {\slshape \bibinfo{booktitle}{Reliable Software Technologies - Ada-Europe 2011 - 16th Ada-Europe International Conference on Reliable Software Technologies, Edinburgh, UK, June 20-24, 2011. Proceedings}}, {\slshape \bibinfo{series}{Lecture Notes in Computer Science}} \bibinfo{volume}{6652}, \bibinfo{publisher}{Springer}, pp. \bibinfo{pages}{211--213}, \doi{10.1007/978-3-642-21338-0\_19}.

\bibitemdeclare{misc}{EUAIAct}
\bibitem{EUAIAct}
\bibinfo{author}{\surnamestart {European Commission}\surnameend} (\bibinfo{year}{2021}): \emph{\bibinfo{title}{Proposal for a Regulation of the European Parliament and of the Council Laying Down Harmonised Rules on Artificial Intelligence (Artificial Intelligence Act)}}.
\newblock \bibinfo{howpublished}{\url{https://artificialintelligenceact.eu/}}.
\newblock \bibinfo{note}{European Commission, COM/2021/206 final}.

\bibitemdeclare{article}{DBLP:journals/fmsd/FalconeMFR11}
\bibitem{DBLP:journals/fmsd/FalconeMFR11}
\bibinfo{author}{Yli{\`{e}}s \surnamestart Falcone\surnameend}, \bibinfo{author}{Laurent \surnamestart Mounier\surnameend}, \bibinfo{author}{Jean{-}Claude \surnamestart Fernandez\surnameend} \& \bibinfo{author}{Jean{-}Luc \surnamestart Richier\surnameend} (\bibinfo{year}{2011}): \emph{\bibinfo{title}{Runtime enforcement monitors: composition, synthesis, and enforcement abilities}}.
\newblock {\slshape \bibinfo{journal}{Formal Methods Syst. Des.}} \bibinfo{volume}{38}(\bibinfo{number}{3}), pp. \bibinfo{pages}{223--262}, \doi{10.1007/S10703-011-0114-4}.

\bibitemdeclare{inproceedings}{DBLP:conf/sefm/FerrandoM22}
\bibitem{DBLP:conf/sefm/FerrandoM22}
\bibinfo{author}{Angelo \surnamestart Ferrando\surnameend} \& \bibinfo{author}{Vadim \surnamestart Malvone\surnameend} (\bibinfo{year}{2022}): \emph{\bibinfo{title}{Runtime Verification with Imperfect Information Through Indistinguishability Relations}}.
\newblock In \bibinfo{editor}{Bernd{-}Holger \surnamestart Schlingloff\surnameend} \& \bibinfo{editor}{Ming \surnamestart Chai\surnameend}, editors: {\slshape \bibinfo{booktitle}{Software Engineering and Formal Methods - 20th International Conference, {SEFM} 2022, Berlin, Germany, September 26-30, 2022, Proceedings}}, {\slshape \bibinfo{series}{Lecture Notes in Computer Science}} \bibinfo{volume}{13550}, \bibinfo{publisher}{Springer}, pp. \bibinfo{pages}{335--351}, \doi{10.1007/978-3-031-17108-6\_21}.

\bibitemdeclare{inproceedings}{DBLP:conf/cav/HiplerKLS24}
\bibitem{DBLP:conf/cav/HiplerKLS24}
\bibinfo{author}{Raik \surnamestart Hipler\surnameend}, \bibinfo{author}{Hannes \surnamestart Kallwies\surnameend}, \bibinfo{author}{Martin \surnamestart Leucker\surnameend} \& \bibinfo{author}{C{\'{e}}sar \surnamestart S{\'{a}}nchez\surnameend} (\bibinfo{year}{2024}): \emph{\bibinfo{title}{General Anticipatory Runtime Verification}}.
\newblock In \bibinfo{editor}{Arie \surnamestart Gurfinkel\surnameend} \& \bibinfo{editor}{Vijay \surnamestart Ganesh\surnameend}, editors: {\slshape \bibinfo{booktitle}{Computer Aided Verification - 36th International Conference, {CAV} 2024, Montreal, QC, Canada, July 24-27, 2024, Proceedings, Part {II}}}, {\slshape \bibinfo{series}{Lecture Notes in Computer Science}} \bibinfo{volume}{14682}, \bibinfo{publisher}{Springer}, pp. \bibinfo{pages}{133--155}, \doi{10.1007/978-3-031-65630-9\_7}.

\bibitemdeclare{article}{DBLP:journals/air/HuangRHJDWBMQZCZWXWFM24}
\bibitem{DBLP:journals/air/HuangRHJDWBMQZCZWXWFM24}
\bibinfo{author}{Xiaowei \surnamestart Huang\surnameend}, \bibinfo{author}{Wenjie \surnamestart Ruan\surnameend}, \bibinfo{author}{Wei \surnamestart Huang\surnameend}, \bibinfo{author}{Gaojie \surnamestart Jin\surnameend}, \bibinfo{author}{Yi~\surnamestart Dong\surnameend}, \bibinfo{author}{Changshun \surnamestart Wu\surnameend}, \bibinfo{author}{Saddek \surnamestart Bensalem\surnameend}, \bibinfo{author}{Ronghui \surnamestart Mu\surnameend}, \bibinfo{author}{Yi~\surnamestart Qi\surnameend}, \bibinfo{author}{Xingyu \surnamestart Zhao\surnameend}, \bibinfo{author}{Kaiwen \surnamestart Cai\surnameend}, \bibinfo{author}{Yanghao \surnamestart Zhang\surnameend}, \bibinfo{author}{Sihao \surnamestart Wu\surnameend}, \bibinfo{author}{Peipei \surnamestart Xu\surnameend}, \bibinfo{author}{Dengyu \surnamestart Wu\surnameend}, \bibinfo{author}{Andr{\'{e}} \surnamestart Freitas\surnameend} \& \bibinfo{author}{Mustafa~A. \surnamestart Mustafa\surnameend} (\bibinfo{year}{2024}): \emph{\bibinfo{title}{A survey of safety and
  trustworthiness of large language models through the lens of verification and validation}}.
\newblock {\slshape \bibinfo{journal}{Artif. Intell. Rev.}} \bibinfo{volume}{57}(\bibinfo{number}{7}), p. \bibinfo{pages}{175}, \doi{10.1007/S10462-024-10824-0}.

\bibitemdeclare{misc}{ISO26262}
\bibitem{ISO26262}
\bibinfo{author}{\surnamestart {International Organization for Standardization}\surnameend} (\bibinfo{year}{2018}): \emph{\bibinfo{title}{ISO 26262: Road vehicles – Functional safety}}.
\newblock \bibinfo{howpublished}{\url{https://www.iso.org/standard/68383.html}}.
\newblock \bibinfo{note}{International Organization for Standardization, first edition 2011, updated 2018}.

\bibitemdeclare{inproceedings}{Levy2025StatisticalRV}
\bibitem{Levy2025StatisticalRV}
\bibinfo{author}{Natan \surnamestart Levy\surnameend}, \bibinfo{author}{Adiel \surnamestart Ashrov\surnameend} \& \bibinfo{author}{Guy \surnamestart Katz\surnameend} (\bibinfo{year}{2025}): \emph{\bibinfo{title}{Statistical Runtime Verification for LLMs via Robustness Estimation}}.
\newblock In \bibinfo{editor}{Bettina \surnamestart K{\"{o}}nighofer\surnameend} \& \bibinfo{editor}{Hazem \surnamestart Torfah\surnameend}, editors: {\slshape \bibinfo{booktitle}{Runtime Verification - 25th International Conference, {RV} 2025, Graz, Austria, September 15-19, 2025, Proceedings}}, {\slshape \bibinfo{series}{Lecture Notes in Computer Science}} \bibinfo{volume}{16087}, \bibinfo{publisher}{Springer}, pp. \bibinfo{pages}{457--476}, \doi{10.1007/978-3-032-05435-7\_25}.

\bibitemdeclare{inproceedings}{DBLP:conf/allerton/LindemannQDP23}
\bibitem{DBLP:conf/allerton/LindemannQDP23}
\bibinfo{author}{Lars \surnamestart Lindemann\surnameend}, \bibinfo{author}{Xin \surnamestart Qin\surnameend}, \bibinfo{author}{Jyotirmoy~V. \surnamestart Deshmukh\surnameend} \& \bibinfo{author}{George~J. \surnamestart Pappas\surnameend} (\bibinfo{year}{2023}): \emph{\bibinfo{title}{Conformal Prediction for {STL} Runtime Verification}}.
\newblock In: {\slshape \bibinfo{booktitle}{59th Annual Allerton Conference on Communication, Control, and Computing, Allerton 2023, Monticello, IL, USA, September 26-29, 2023}}, \bibinfo{publisher}{{IEEE}}, p.~\bibinfo{pages}{1}, \doi{10.1109/ALLERTON58177.2023.10313399}.

\bibitemdeclare{article}{DBLP:journals/corr/abs-2502-18917}
\bibitem{DBLP:journals/corr/abs-2502-18917}
\bibinfo{author}{Chuyue \surnamestart Sun\surnameend}, \bibinfo{author}{Viraj \surnamestart Agashe\surnameend}, \bibinfo{author}{Saikat \surnamestart Chakraborty\surnameend}, \bibinfo{author}{Jubi \surnamestart Taneja\surnameend}, \bibinfo{author}{Clark~W. \surnamestart Barrett\surnameend}, \bibinfo{author}{David~L. \surnamestart Dill\surnameend}, \bibinfo{author}{Xiaokang \surnamestart Qiu\surnameend} \& \bibinfo{author}{Shuvendu~K. \surnamestart Lahiri\surnameend} (\bibinfo{year}{2025}): \emph{\bibinfo{title}{ClassInvGen: Class Invariant Synthesis using Large Language Models}}.
\newblock {\slshape \bibinfo{journal}{CoRR}} \bibinfo{volume}{abs/2502.18917}, \doi{10.48550/ARXIV.2502.18917}.
\newblock \eprint{2502.18917}.

\bibitemdeclare{misc}{wang2025pro2guardproactiveruntimeenforcement}
\bibitem{wang2025pro2guardproactiveruntimeenforcement}
\bibinfo{author}{Haoyu \surnamestart Wang\surnameend}, \bibinfo{author}{Chris~M. \surnamestart Poskitt\surnameend}, \bibinfo{author}{Jun \surnamestart Sun\surnameend} \& \bibinfo{author}{Jiali \surnamestart Wei\surnameend} (\bibinfo{year}{2025}): \emph{\bibinfo{title}{Pro2Guard: Proactive Runtime Enforcement of LLM Agent Safety via Probabilistic Model Checking}}.
\newblock \eprint{2508.00500}.

\bibitemdeclare{inproceedings}{DBLP:conf/iclr/0001BN24}
\bibitem{DBLP:conf/iclr/0001BN24}
\bibinfo{author}{Haoze \surnamestart Wu\surnameend}, \bibinfo{author}{Clark~W. \surnamestart Barrett\surnameend} \& \bibinfo{author}{Nina \surnamestart Narodytska\surnameend} (\bibinfo{year}{2024}): \emph{\bibinfo{title}{Lemur: Integrating Large Language Models in Automated Program Verification}}.
\newblock In: {\slshape \bibinfo{booktitle}{The Twelfth International Conference on Learning Representations, {ICLR} 2024, Vienna, Austria, May 7-11, 2024}}, \bibinfo{publisher}{OpenReview.net}.
\newblock \urlprefix\url{https://openreview.net/forum?id=Q3YaCghZNt}.

\bibitemdeclare{inproceedings}{DBLP:conf/nfm/ZhangLD12}
\bibitem{DBLP:conf/nfm/ZhangLD12}
\bibinfo{author}{Xian \surnamestart Zhang\surnameend}, \bibinfo{author}{Martin \surnamestart Leucker\surnameend} \& \bibinfo{author}{Wei \surnamestart Dong\surnameend} (\bibinfo{year}{2012}): \emph{\bibinfo{title}{Runtime Verification with Predictive Semantics}}.
\newblock In \bibinfo{editor}{Alwyn \surnamestart Goodloe\surnameend} \& \bibinfo{editor}{Suzette \surnamestart Person\surnameend}, editors: {\slshape \bibinfo{booktitle}{{NASA} Formal Methods - 4th International Symposium, {NFM} 2012, Norfolk, VA, USA, April 3-5, 2012. Proceedings}}, {\slshape \bibinfo{series}{Lecture Notes in Computer Science}} \bibinfo{volume}{7226}, \bibinfo{publisher}{Springer}, pp. \bibinfo{pages}{418--432}, \doi{10.1007/978-3-642-28891-3\_37}.

\bibitemdeclare{article}{DBLP:journals/corr/abs-2412-06512}
\bibitem{DBLP:journals/corr/abs-2412-06512}
\bibinfo{author}{Yedi \surnamestart Zhang\surnameend}, \bibinfo{author}{Yufan \surnamestart Cai\surnameend}, \bibinfo{author}{Xinyue \surnamestart Zuo\surnameend}, \bibinfo{author}{Xiaokun \surnamestart Luan\surnameend}, \bibinfo{author}{Kailong \surnamestart Wang\surnameend}, \bibinfo{author}{Zhe \surnamestart Hou\surnameend}, \bibinfo{author}{Yifan \surnamestart Zhang\surnameend}, \bibinfo{author}{Zhiyuan \surnamestart Wei\surnameend}, \bibinfo{author}{Meng \surnamestart Sun\surnameend}, \bibinfo{author}{Jun \surnamestart Sun\surnameend}, \bibinfo{author}{Jing \surnamestart Sun\surnameend} \& \bibinfo{author}{Jin~Song \surnamestart Dong\surnameend} (\bibinfo{year}{2024}): \emph{\bibinfo{title}{The Fusion of Large Language Models and Formal Methods for Trustworthy {AI} Agents: {A} Roadmap}}.
\newblock {\slshape \bibinfo{journal}{CoRR}} \bibinfo{volume}{abs/2412.06512}, \doi{10.48550/ARXIV.2412.06512}.
\newblock \eprint{2412.06512}.

\bibitemdeclare{article}{DBLP:journals/corr/abs-2505-18585}
\bibitem{DBLP:journals/corr/abs-2505-18585}
\bibinfo{author}{Yedi \surnamestart Zhang\surnameend}, \bibinfo{author}{Sun~Yi \surnamestart Emma\surnameend}, \bibinfo{author}{Annabelle Lee~Jia \surnamestart En\surnameend} \& \bibinfo{author}{Jin~Song \surnamestart Dong\surnameend} (\bibinfo{year}{2025}): \emph{\bibinfo{title}{RvLLM: {LLM} Runtime Verification with Domain Knowledge}}.
\newblock {\slshape \bibinfo{journal}{CoRR}} \bibinfo{volume}{abs/2505.18585}, \doi{10.48550/ARXIV.2505.18585}.
\newblock \eprint{2505.18585}.

\bibitemdeclare{article}{DBLP:journals/corr/abs-2504-02964}
\bibitem{DBLP:journals/corr/abs-2504-02964}
\bibinfo{author}{Yiqi \surnamestart Zhao\surnameend}, \bibinfo{author}{Emily \surnamestart Zhu\surnameend}, \bibinfo{author}{Bardh \surnamestart Hoxha\surnameend}, \bibinfo{author}{Georgios \surnamestart Fainekos\surnameend}, \bibinfo{author}{Jyotirmoy \surnamestart Deshmukh\surnameend} \& \bibinfo{author}{Lars \surnamestart Lindemann\surnameend} (\bibinfo{year}{2025}): \emph{\bibinfo{title}{Distributionally Robust Predictive Runtime Verification under Spatio-Temporal Logic Specifications}}.
\newblock {\slshape \bibinfo{journal}{ACM Trans. Cyber-Phys. Syst.}}, \doi{10.1145/3748818}.

\end{thebibliography}

\end{document}